\documentclass[12pt]{article}

\RequirePackage[OT1]{fontenc}
\RequirePackage{amsthm,amsmath}
\RequirePackage[authoryear]{natbib}
\RequirePackage[colorlinks,citecolor=blue,urlcolor=blue]{hyperref}
\RequirePackage{hypernat}
\usepackage{multirow}
\usepackage[small]{caption}
\usepackage{bm}
\usepackage[hmargin=1.2in, vmargin=1.2in]{geometry}
\usepackage{amssymb}
\usepackage{graphicx}
\usepackage{lineno}
  
\usepackage{setspace}
\numberwithin{equation}{section} 
\newtheorem{thm}{Theorem}

\newtheorem*{doob}{Doob's Theorem}

\theoremstyle{remark}

\newcommand{\inv}[1]{#1^{-1}}

\newcommand{\ind}[1]{{\mathbf{1}}{\left(#1\right)}}

\def\id{\perp\!\!\!\perp}
 
\DeclareMathOperator{\diag}{diag}
\DeclareMathOperator{\tr}{tr}

\newcommand{\sumi}[1]{\sum_{h = 1}^{\MakeLowercase{#1}}}
\newcommand{\beq}{\begin{equation}}
\newcommand{\eeq}{\end{equation}}

\newcommand{\ben}{\begin{enumerate}}
\newcommand{\een}{\end{enumerate}}

\def\wpone{ \overset {as}{\longrightarrow}}

\newcommand{\keywords}[1]{\par\noindent 
{\small{\em Keywords\/}: #1}}

\makeatletter\def\blfootnote{\xdef\@thefnmark{}\@footnotetext}\makeatother
\begin{document}


\begin{center}
\LARGE{Bayesian Gaussian Copula Factor Models for Mixed Data}
\end{center}
 \thispagestyle{empty}

\large{Jared S. Murray\quad
 David B. Dunson\quad
 Lawrence Carin\quad
 Joseph E. Lucas}
\blfootnote{\noindent Jared S. Murray is PhD Student, Dept. of Statistical Science, Duke University, Durham, NC 27708 (E-mail: jared.murray@stat.duke.edu). David B. Dunson is Professor, Dept. of Statistical Science, Duke University  Durham, NC 27708 (E-mail: dunson@stat.duke.edu). 
Lawrence Carin is William H. Younger Professor, Dept. of Electrical \& Computer Engineering, Pratt School of Engineering, Duke University,Durham, NC 27708 (E-mail: lcarin@ee.duke.edu)
Joseph E. Lucas is Assistant Research Professor, Institute for Genome Sciences and Policy, Duke University, Durham, NC 27710 (E-mail: joe@stat.duke.edu)
We would like to acknowledge the support of the Measurement to Understand Re-Classifcation of Disease of Cabarrus and Kannapolis (MURDOCK) Study and the NIH CTSA (Clinical and Translational Science Award) 1UL1RR024128-01, without whom this research would not have been possible. The first author would also like to thank Richard Hahn, Jerry Reiter and Scott Schmidler for helpful discussion.}

\normalsize

\begin{abstract}
Gaussian factor models have proven widely useful for parsimoniously characterizing dependence in multivariate data.  There is a rich literature on their extension to mixed categorical and continuous variables, using latent Gaussian variables or through generalized latent trait models acommodating measurements in the exponential family. However, when generalizing to non-Gaussian measured variables the latent variables typically influence both the dependence structure and the form of the marginal distributions, complicating interpretation and introducing artifacts.  To address this problem we propose a novel class of Bayesian Gaussian copula factor models which decouple the latent factors from the marginal distributions.  A semiparametric specification for the marginals based on the extended rank likelihood yields straightforward implementation and substantial computational gains.  We provide new theoretical and empirical justifications for using this likelihood in Bayesian inference. We propose new default priors for the factor loadings and develop efficient parameter-expanded Gibbs sampling for posterior computation.  The methods are evaluated through simulations and applied to a dataset in political science.  The models in this paper are implemented in the R package bfa.
\end{abstract}

\section{Introduction}\label{section:intro2}
\setcounter{page}{1}
Factor analysis and its generalizations are powerful tools for analyzing and exploring multivariate data, routinely used in applications as diverse as social science, genomics and finance. The typical Gaussian factor model is given by
\beq
\bm y_i = \bm{\Lambda\eta}_i + \bm \epsilon_i\label{eq:fm}
\eeq
where $\bm y_i$ is a $p\times 1$ vector of observed variables, $\bm\Lambda$ is a $p\times k$ matrix of factor loadings ($k<p$), $\bm\eta_i\sim N(\bm 0, \bm I)$ is a $k\times 1$ vector of latent variables or factor scores, and $\bm\epsilon_i\sim N(\bm 0, \bm\Sigma)$ are idiosyncratic noise with $\bm\Sigma = \diag(\sigma^2_1, \dots, \sigma_p^2)$.  Marginalizing out the latent variables, $\bm y_i\sim N(\bm 0, \bm{\Lambda\Lambda}'+\bm\Sigma)$, so that the covariance in $\bm y_i$ is explained by the (lower-dimensional) latent factors.  The model in \eqref{eq:fm} may be generalized to incorporate covariates at the level of the observed or latent variables, or to allow dependence between the latent factors. For exposition we focus on this simple case.

This model has been extended to data with mixed measurement scales, often by linking observed categorical variables to underlying Gaussian variables which follow a latent factor model (e.g.  \cite{Muthen1983a}).
An alternative is to include shared latent factors in separate generalized linear models for each observed variable  
\citep{Sammel1997a, Moustaki2000, Dunson2003, Dunson2000}. Unlike in the Gaussian factor model the latent variables typically impact both dependence and the form of the marginal distributions.  For example, suppose $y_i = (y_{i1},y_{i2})'$ are bivariate counts assigned Poisson log-linear models: 
$\log \mbox{E}(y_{ij}\, |\, \eta_i) = \mu_j + \lambda \eta_i$.
Then $\lambda$ governs both the dependence between $y_{i1},\ y_{i2}$ and the overdispersion in each marginal distribution.  This confounding can lead to substantial artifacts and misleading inferences. Additionally, computation in such models is difficult and requiring marginal distributions in the exponential family can be restrictive. 

There is a growing literature on semiparametric latent factor models to address the latter problem.  A number of authors have proposed mixtures of factor models \citep{Ghahramani2000,  Chen2010}. \cite{Song2010b} instead allow flexible error distributions in Eq. \eqref{eq:fm}. \cite{Yang2010a} proposed a broad class of semiparametric  structural equation models which allow an unknown distribution for $\bm \eta_i$.  When building such flexible mixture models there is a sacrifice to be made in terms of interpretation, parsimony and computation, and subtle confounding effects remain. It would be appealing to retain the simplicity, interpretability and computational scalability of Gaussian factor models while allowing the marginal distributions to be unknown and free of the dependence structure.  

To accomplish these ambitious goals we propose a semiparametric Bayesian Gaussian copula model utilizing the extended rank likelihood of \cite{hoff-erl} for the marginal distributions. This approximation avoids a full model specification and is in some sense not fully Bayesian, but in practice we expect that this rank-based likelihood discards only a mild amount of information while providing robust inference. An additional contribution of this paper is to provide new theoretical and empirical justification for this approach.

The paper proceeds as follows: In Section 2, we propose the Gaussian copula factor model for mixed scale data and discuss its relationship to existing models.  In Section 3 we develop a Bayesian approach to inference, specifying prior distributions and outlining a straightforward and efficient Gibbs sampler for posterior computation.  Section 4 contains a simulation study, and Section 5 illustrates the utility of our method in a political science application. Section 6 concludes with a discussion.

\section{The Gaussian copula factor model}\label{section:gausscop}

A $p$-dimensional copula $\mathbb{C}$ is a distribution function on $[0,1]^p$ where each univariate marginal distribution is uniform on $[0,1]$. Any joint distribution $F$ can be completely specified by its marginal distributions and a copula; that is, there exists a copula $\mathbb{C}$ such that
\begin{equation}
F(y_1, \dots, y_p) = \mathbb{C}(F_1(y_1), \dots, F_p(y_p))\label{eq:cop}
\end{equation}
where $F_j$ are the univariate marginal distributions of $F$ \citep{sklar}. If all $F_j$ are continuous then $\mathbb{C}$ is unique, otherwise it is uniquely determined on $\text{Ran}(F_1)\times\dots\times\text{Ran}(F_p)$ where $\text{Ran}(F_j)$ is the range of $F_j$. The copula of a multivariate distribution encodes its dependence structure, and is invariant to strictly increasing transformations of its univariate margins. Here we consider the Gaussian copula:
\begin{equation}
\mathbb{C}(u_1, \dots u_p) = \Phi_p(\inv \Phi(u_1), \dots, \inv \Phi(u_p)\ |\ \bm C),\quad (u_1, \dots, u_p)\in [0,1]^p\label{eq:gaussiancop}
\end{equation}
where $\Phi_p(\cdot | \bm C)$ is the $p$-dimensional Gaussian cdf with correlation matrix $\bm C$ and $\Phi$ is the univariate standard normal cdf. Combining \eqref{eq:cop} and \eqref{eq:gaussiancop} we have 
\begin{equation}
F(y_1, \dots, y_p) = \Phi_p(\Phi^{-1}(F_1(y_1)), \dots, \Phi^{-1}(F_p(y_p))\ |\ \bm C)\label{eq:gcopdist}
\end{equation}
From \eqref{eq:gcopdist} a number of properties are clear: The joint marginal distribution of any subset of $\bm y$ has a Gaussian copula with correlation matrix given by the appropriate submatrix of $\bm C$, and $y_{j}\id y_{j'}$ if and only if $c_{jj'}=0$. When $F_j,\ F_{j'}$ are continuous, $c_{jj'} = Corr(\Phi^{-1}(F_j(y_{j}), \Phi^{-1}(F_{j'}(y_{j'}))$ which is an upper bound on $Corr(y_{j},y_{j'})$ (attained when the margins are Gaussian) \citep{ Klaassen1997}. The rank correlation coefficients Kendall's tau and Spearman's rho are monotonic functions of $c_{jj'}$ \citep{Hult2002}. For variables taking finitely many values $c_{jj'}$ gives the polychoric correlation coefficient \citep{Muthen1983a}.

If the margins are all continuous then zeros in $\bm R = \bm C^{-1}$ imply conditional independence, as in the multivariate Gaussian distribution. However this is generally not the case when some variables are discrete. Even in the simple case where $p=3$, $Y_3$ is discrete and $c_{13}c_{23}\neq 0$, if $r_{12}=0$ then $Y_1$ and $Y_2$ are in fact \emph{dependent} conditional on $Y_3$ (a similar result holds when conditioning on several continuous variables and a discrete variable as well - details available in supplementary materials). Results like these suggest that sparsity priors for $\bm R$ in Gaussian copula models (e.g. \cite{pitt2006, Dobra2011}) are perhaps not always well-motivated when discrete variables are present, and should be interpreted only with great care.

A Gaussian copula model can be expressed in terms of latent Gaussian variables $\bm z$:   Let $F_j^{-1}(t) = \inf\{ t : F_j(y) \geq t, y\in \mathbb{R}\} $ be the usual pseudo-inverse of $F_j$ and suppose $\bm \Omega$ is a covariance matrix with $\bm C$ as its correlation matrix.
If $\bm z\sim N(0, \bm\Omega)$ and $y_j = \inv F_j(\Phi(z_j/\sqrt{\omega_{jj}}))$ for $1\leq j\leq p$ then $F(\bm y)$ has a Gaussian copula with correlation matrix $\bm C$ and univariate marginals $F_j$. We utilize this representation to generalize the Gaussian factor model to Gaussian copula factor models by assigning $\bm z$ a latent factor model:
\[
\bm\eta_i\sim N(0, \bm I),\quad \bm z_i|\bm\eta_i\sim N(
\bm\Lambda\bm\eta_i,\ \bm I)
\]
\beq
y_{ij} = \inv F_j\left(\Phi\left(\frac{z_{ij}}{\sqrt{1+\sum_{h=1}^k \lambda_{jh}^2}}\right)\right)\label{eq:gcfm}
\eeq
Inference takes place on the scaled loadings
\beq
\tilde\lambda_{jh} = \frac{\lambda_{jh}}{\sqrt{1+\sumi{k} \lambda_{jh}^2}}
\eeq
so that $c_{jj'} = \sumi{k}\tilde\lambda_{jh}\tilde\lambda_{j'h}$. Rescaling is important as the factor loadings are not otherwise comparable across the different variables - even though $\bm\Lambda$ is technically identified it is not easily interpreted.  We also consider the \emph{uniqueness} of variable $j$, given by
\beq
u_j = 1 - \sumi{k}\tilde\lambda_{jh}^2 = \frac{1}{1 + \sumi{k}\lambda_{jh}^2}\label{eq:uniqueness}
\eeq
In the Gaussian factor model $u_j$ is $\sigma^2_j/(\sigma^2_j + \sumi{k}\lambda^2_{jh})$, the proportion of variance unexplained by the latent factors. In the Gaussian \emph{copula} factor model this exact interpretation does not hold, but $u_j$ still represents a measure of dependence on common factors. 

\subsection{Relationship to existing factor models}\label{section:existing}

The Gaussian factor model and probit factor models are both special cases of the Gaussian copula factor model. Probit factor models for binary or ordered categorical data parameterize each  margin by a collection of ``cutpoints" $\gamma_{j0}, \dots\gamma_{jc_j}$ (taking $\gamma_{j0}=-\infty$ and $\gamma_{jc_j}=\infty$ without loss of generality) so that
$F_j(c) = \Phi\left(\gamma_{jc}(1+\sum_{h=1}^k \lambda_{jh}^2)^{-1/2}\right)$. Then $F_j$ has the pseudoinverse 
\[
\inv F (u_{ij}) = \sum_{c=1}^{c_j} c
\ind{\Phi\left(\frac{\gamma_{jc-1}}{\sqrt{1+\sum_{h=1}^k \lambda_{jh}^2}}\right) < u_{ij} \leq \Phi\left(\frac{\gamma_{jc}}{\sqrt{1+\sum_{h=1}^k \lambda_{jh}^2}}\right) }
\]
Plugging this into \eqref{eq:gcfm} and simplifying gives $y_{ij}= \sum_{c=1}^{C_j} c\ind{ \gamma_{jc-1} < z_{ij} \leq \gamma_{jc}}$
where $\bm z_i\sim N(\bm 0, \bm{\Lambda\Lambda}' + \bm I)$, the data augmented representation of an ordinal probit factor model. Naturally the connection extends to mixed Gaussian/probit margins as well.

Other factor models which have Gaussian/probit models as special cases include semiparametric factor models, which assume non-Gaussian latent variables $\bm \eta_i$ or errors $\bm \epsilon_i$, retaining the linear model formulation \eqref{eq:fm} so that marginally $Cov(\bm y_i) = \bm{\Lambda} Cov(\bm\eta_i)\bm{\Lambda}' + \bm\Sigma$. But $F(\bm y_i)$ no longer has a Gaussian copula, and since the joint distribution is no longer elliptically symmetric the covariance matrix is unlikely to be an adequate measure of dependence. Further, the dependence and marginal distributions are confounded since the implied correlation matrix will depend on the parameters of the marginal distributions.

Our model overcomes these shortcomings. In the Gaussian copula factor model $\bm{\tilde\Lambda}$ governs the dependence separately from the marginal distributions, representing a factor-analytic decomposition for the scale-free copula parameter $\bm C$ rather than $Cov(\bm y_i)$.  The Gaussian copula model is also invariant to strictly monotone transformations of univariate margins. Therefore it is consistent with the common assumption that there exist monotonic functions $h_1,\dots, h_p$ such that $(h_1(y_1),\dots h_p(y_p))'$ follows a Gaussian factor model, while existing semiparametric approaches are not. Researchers using our method are  not required to consider numerous univariate transformations to achieve ``approximate normality".

\subsection{Marginal Distributions}

One way to deal with marginal distributions in a copula model is to specify a parametric family for each margin and infer the parameters simultaneously with $\bm C$ (see e.g. \cite{pitt2006} for a Bayesian approach). This is computationally expensive for even moderate $p$, and there is often no obvious choice of parametric family for every margin. Since our goal is not to learn the whole joint distribution but rather to characterize its dependence structure we would prefer to treat the marginal distributions as nuisance parameters. 

When the data are all continuous a popular semiparametric method is a two-stage approach in which an estimator $\hat F_j$ is used to compute ``pseudodata" $\hat z_{ij} = \inv\Phi({\hat{F_j}}(y_{ij}))$, which are treated as fixed to infer the copula parameters. A natural candidate is $\hat F_j(t) = \frac{n}{n+1}\sum_{i=1}^n \frac{1}{n}\ind{y_{ij}\leq t}$, the (scaled) empirical marginal cdf. \cite{ Klaassen1997} considered such estimators in the Gaussian copula and \cite{Genest1995} developed them in the general case. 
 However, this method cannot handle discrete margins. To accommodate mixed discrete and continuous data \cite{hoff-erl} proposed an approximation to the full likelihood called the extended rank likelihood, derived as follows: Since the transformation $y_{ij} = \inv F_j( \Phi(z_{ij}))$ is nondecreasing, when we observe $\bm y_j = (y_{1j}, \dots, y_{nj})$ we also observe a partial ordering on $\bm z_{j} = (z_{1j}, \dots z_{nj})$. To be precise we have that 
\begin{equation}
\bm z_{j} \in D(\bm y_{j}) \equiv  \{ \bm z_{j} \in \mathbb{R}^n : y_{ij} < y_{i'j} \Rightarrow  z_{ij} < z_{i'j}   \} \label{erlset} 
\end{equation}
The set $D(\bm y_{j})$ is just the set of possible $\bm z_{j}=(z_{1j},\dots, z_{nj})$ which are consistent with the ordering of the observed data on the $ j^{th}$ variable. 
Let ${D}(\bm Y) = \{ \bm Z\in \mathbb{R}^{p\times n} : \bm z_j \in D(\bm y_j)\quad \forall\ 1\leq j\leq p \}$. 
Then we have
\begin{align}
  P(\bm Y | \bm C, F_1, \dots F_p ) &= P(\bm Y, \bm Z\in {D}(\bm Y) | \bm C, F_1, \dots F_p)\nonumber \\
  &=P(\bm Z\in {D}(\bm Y) | \bm C) \times P(\bm Y| \bm Z\in {D}(\bm Y) , \bm C, F_1, \dots F_p)\label{suff}
  \end{align}
  where \eqref{suff} holds because given $\bm C$ the event $\bm Z\in {D}(\bm Y)$ does not depend on the marginal distributions. \cite{hoff-erl} proposes dropping the second term in \eqref{suff} and using $P(\bm Z\in {D}(\bm Y) | \bm C)$ as the likelihood. Intuitively we would expect the first term to include most of the information about $\bm C$. Simulations in Section \ref{section:simstudy} provide further evidence of this. \cite{hoff-erl} shows that when the margins are all continuous the marginal ranks satisfy certain relaxed notions of sufficiency for $\bm C$, although these fail when some margins are discrete. 
 Unfortunately theoretical results for applications involving mixed data have been lacking.
  
  To address this we give a new proof of strong posterior consistency for $\bm C$ under the extended rank likelihood with nearly any mixture of discrete and continuous margins (barring pathological cases which preclude identification of $\bm C$). Posterior consistency will generally fail under Gaussian/probit models when any margin is misspecified. A similar result for continuous data and a univariate rank likelihood was obtained by \cite{Gu2009}. We replace $\bm Y$ with $\bm  Y^{(n)}$ for notational clarity below. 

\begin{thm}
Let $\Pi$ be a prior distribution on the space of all positive semidefinite correlation matrices $\mathcal{C}$ with corresponding density $\pi (\bm C)$ with respect to a measure $\nu$. Suppose $\pi(\bm{C})>0$ almost everywhere with respect to $\nu$ and that $F_1, \dots, F_p$,  are the true marginal cdfs. Then for $\bm C_0$ a.e. $[\nu]$ and any neighborhood $\mathcal{A}$ of $\bm C_0$ we have that
\beq
\lim_{n\rightarrow \infty} \Pi(\bm C\in \mathcal{A}\ |\ \bm Z^{(n)}\in D(\bm Y^{(n)})) = 1\ \text{a.s.}\ [G^\infty_{\bm C_0, F_1, \dots, F_p}]\label{}
\eeq
where $G^\infty_{\bm C_0, F_1, \dots, F_p}$ is the distribution of $\{ \bm y_i\}_{i=1}^\infty$ under $\bm C_0, F_1, \dots F_p$. 
\end{thm}

The proof is in Appendix \ref{appendix:proof}. 
We assumed a prior $\pi(\bm C)$ having full support on $\mathcal{C}$. Under factor-analytic priors fixing $k<p$ restricts the support of $\pi$, and posterior consistency will only hold if $\bm C_0$ has a factor analytic decomposition in $k$ or fewer factors. But by setting $k$ large (or inferring it) it is straightforward to define factor-analytic priors which have full-support on $\mathcal{C}$ (further discussion in Section \ref{section:conclusion}). In practice, many correlation matrices which do not exactly have a $k$-factor decomposition are still well-approximated by a $k$-factor model. The result also applies to posterior consistency for $\bm{\tilde\Lambda}$ if $k$ is chosen correctly, given compatible identifying restrictions. Theorem 1 also implies that for two divergent priors on $\Lambda$, the predictive distributions of $\bm z_i$ and $\eta_i$ will still converge, which is perhaps more directly relevant when e.g. estimating factor scores as in Section \ref{section:perisk}.

The efficiency of semiparametric estimators such as ours is also an important issue. \cite{Hoff2011} give some preliminary results which suggest that pseudo-MLE's based on the rank likelihood for continuous margins may be asymptotically relatively efficient. However, it is unclear whether even these results apply to the case of mixed continuous and discrete margins, which is our primary focus. Simulations of the efficiency of posterior means under the extended rank likelihood versus correctly specified parametric models appear in Section \ref{section:sim-eff}. These results give an indication of the worst-case scenario in terms of efficiency lost in using the likelihood approximation, and are quite favorable in general.

\section{Prior Specification and Posterior Inference}\label{section:prior} 
\subsection{Prior Specification}

Since the factor model is invariant under rotation or scaling of the loadings and scores we assume that sufficient identifying conditions are imposed (by introducing sign constraints and fixed zeros in $\bm\Lambda$), or that inference is on $\bm C$ which does not suffer from this indeterminacy. For brevity we also assume here that $k$ is known and fixed. Suggestions for incorporating uncertainty in $k$ are in the Discussion.

A common prior for the unrestricted factor loadings in Gaussian, probit or mixed factor models is $\lambda_{jh}\sim N(0,1/b)$. However, these priors have some troubling properties outside the Gaussian factor model: When  $\sigma_j\equiv 1$ as in probit or mixed Gaussian/probit factor models -- or in our copula model -- the implied prior on $u_j$ is
\beq
\pi(u_j) = \frac{\left(b/2\right)^{-k/2}}{\Gamma(k/2)}\left(\frac{1}{u_j^2}\right)\left(\frac{1-u_j}{u_j}\right)^{k/2-1} \exp\left[-\frac{b}{2}\left(\frac{1-u_j}{u_j}\right)\right]\label{eq:normuniq}
\eeq

\begin{figure}
\begin{center}
\includegraphics[width=\textwidth]{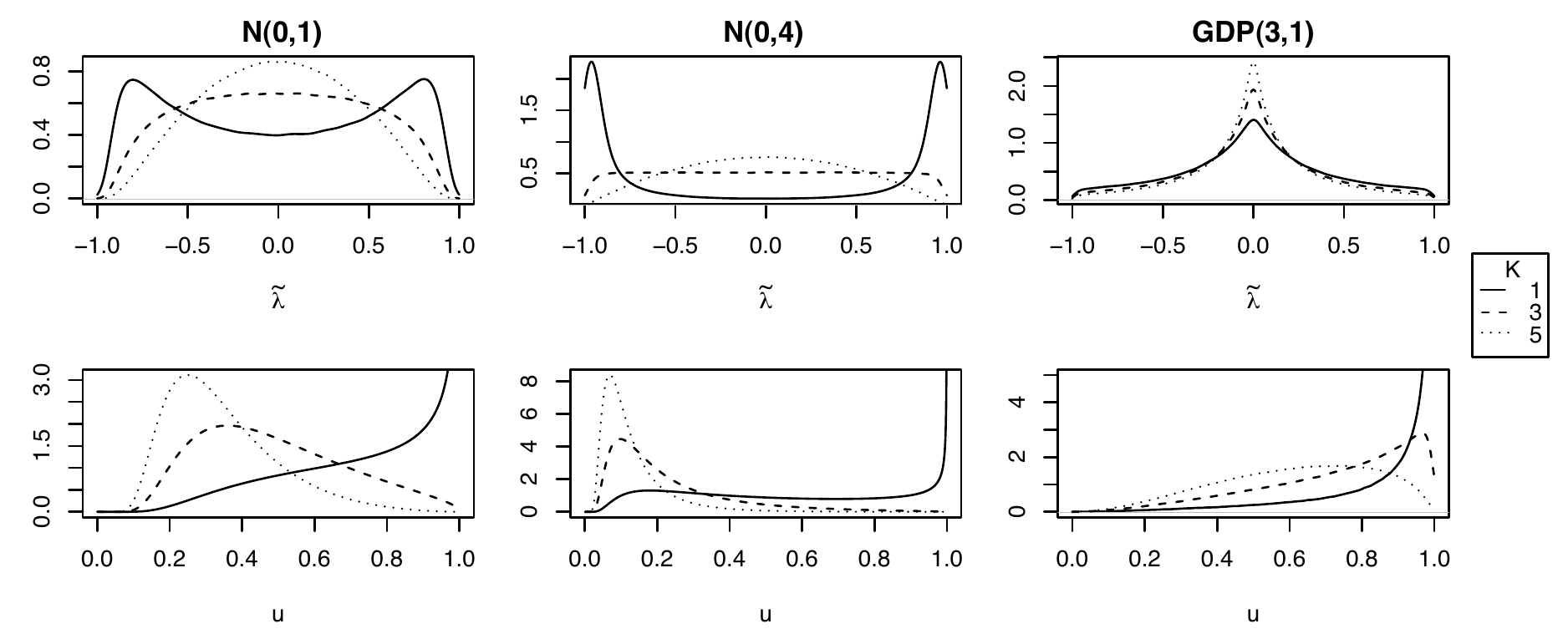}
\end{center}
\vspace{-20pt}
\caption{Induced priors on the scaled factor loadings (top row) and uniquenesses (bottom row) implied by different priors as $K$ varies}
\label{fig:normuniq}
\end{figure}

Figure \ref{fig:normuniq} shows that these priors are quite informative on the uniquenesses, especially as $k$ increases. When $k$ is small they are particularly informative on the scaled loadings, shrinking $\tilde\lambda_{jh}$ toward \emph{large} values, rather than toward zero. This effect becomes worse as the prior variance increases. The problem is that the normal prior puts insufficient mass near zero. Coupled with the normalization this results in a ``smearing" of mass across the columns of $\tilde{\bm\Lambda}$, deflating $u_j$,  inducing spurious correlations, and giving inappropriately high probability to values of the scaled loadings near $\pm 1$. Therefore the normal prior is a very poor default choice in these models.

To address these shortcomings we consider shrinkage priors on $\lambda_{jh}$ which place significant mass at or near zero. Such parsimony is also desirable for more interpretable results. Shrinkage priors have been thoroughly explored in the regression context (see e.g. \cite{Polson2010} and references therein). In that context heavy-tailed distributions are desirable. While somewhat heavy tails are appealing here (so that $\pi(\tilde\lambda_{jh})$ decays slowly to zero as $|\tilde\lambda_{jh}|\rightarrow 1$), \emph{extremely} heavy tails are inappropriate. Very heavy tails imply that with significant probability a single unscaled loading (say $\lambda_{jm}$) in a row $j$ will be much larger than the others so that $\tilde\lambda_{jh}\approx \lambda_{jh}/|\lambda_{jm}|$ for $1\leq h\leq k$. The resulting joint prior on $\bm{\tilde\lambda_j}=(\tilde \lambda_{j1}, \dots, \tilde \lambda_{jk})$ will assign undesirably high probability to vectors with one entry near $\pm 1$ and the rest near $0$, yielding correlations which are approximately $0$ or $\pm 1$. Applying these priors in this new setting requires extra care.

As a default choice we recommend the generalized double Pareto (GDP) prior of \citep{artin-dp} which has the density
\beq
\pi(\lambda_{jh}) = \frac{\alpha}{2\beta}\left( 1+ \frac{|\lambda_{jh}|}{\beta} \right)^{-(\alpha+1)}
\eeq
which we will refer to as $\lambda_{jh}\sim GDP(\alpha, \beta)$.
The GDP is a flexible generalization of the Laplace distribution with a sharper peak at zero and heavier tails. It has the following scale-mixture representation: $\lambda_{jh}|\psi_{jh}\sim N(0, \psi_{jh})$, $\psi_{jh} |\xi_{jh}\sim Exp(\xi_{jh}^2/2)$, and $\xi_{jh}\sim Ga(\alpha, \beta)$ which leads to conditional conjugacy and a simple Gibbs sampler.
The GDP's tail behavior is determined by $\alpha$, and $\beta$ is a scale parameter. \cite{artin-dp} handle the hyperparameters by either fixing them both at $1$ or assigning them a hyperprior. Here taking $\alpha=3$, $\beta=1$ is a good default choice: The $GDP(3,1)$ distribution has mean $0$ and variance $1$, and $Pr(|\lambda_{jh}|<2)\approx 0.96$. Critically, taking $\alpha=3$ leads to tails of $\pi(\lambda_{jh})$ light enough to induce a sensible prior on $\tilde\lambda_{jk}$. 

Figure \ref{fig:normuniq} shows draws from the implied prior on $u_j$ and $\tilde \lambda_{jh}$ under the $GDP(3,1)$ prior, which are much more reasonable than the current default Normal priors. Note that as $K$ increases, the prior puts increasing mass near zero without changing a great deal in the tails. This is reasonable since we expect each variable to load highly on only a few factors, and is difficult to mimic with the light-tailed normal priors. The prior on the uniquenesses remains relatively flat under the GDP prior, while the normal prior increasingly favors lower values and less parsimony.



\subsection{Parameter-Expanded Gibbs Sampling}\label{section:gibbs}

For efficient MCMC inference we introduce a parameter-expanded (PX) version of our original model. The PX approach \citep{Meng1999, Liu1999} adds redundant (non-identified) parameters to reduce serial dependence in MCMC and improve convergence and mixing behavior. Naive Gibbs sampling in our model  suffers from high autocorrelation due to strong dependence between $\bm Z$ and $\bm\Lambda$. We modify \eqref{eq:gcfm} by adding scale parameters $\bm V = \diag(v^2_1,\dots,v^2_p)$ to weaken this dependence:
\begin{align}
\bm w_i &\sim N\left(\bm V^{1/2}\bm \Lambda\bm \eta_i, \bm V\right)\nonumber\\
y_{ij} &= F_j^{-1}\left(\Phi\left(\frac{w_{ij}}{v_j\sqrt{1+\sum_{j=1}^K \lambda_{jh}^2}}\right)\right)\label{eq:pxmodel}
\end{align}
Since $w_{ij}/v_j$ and $z_{ij}$ are equal in distribution \eqref{eq:pxmodel} is observationally equivalent to the original model. We assume that $\bm V$ is independent of the inferential parameters \emph{a priori} so that $\pi(\bm{\Lambda, H, V}|\bm Y)=\pi(\bm {\Lambda, H} |\bm Y) \pi(\bm V)$ (where $\bm H'$ is the $n\times k$ matrix with entries $\eta_{ik}$) and the marginal posterior distribution of the inferential parameters is unchanged.

We choose the conjugate PX prior $1/v_j^2\sim Ga(n_0/2, n_0/2)$ (independently). The greatest benefits from PX are realized when the PX prior is most diffuse, which would imply sending $n_0\rightarrow 0$ and an improper PX prior. The resulting posterior for $(\bm \Lambda, \bm H, \bm V)$ is also improper, but we can prove that the samples of $(\bm \Lambda, \bm H)$ from the corresponding Gibbs sampler still have the desired stationary distribution $\pi(\bm \Lambda, \bm H | \bm Y)$ (Appendix \ref{appendix:px}). The PX Gibbs sampler is implemented as follows:

\textbf{PX parameters:} Draw
$1/v_j^2 \sim Ga(n/2, s_j/2)$ where $\bm\Psi_j = \diag(\psi^2_{j1}/2, \dots, \psi^2_{jh_{j}}/2)$ and
$s_j=\bm{z_j}(\bm I - \bm H_j'(\bm\Psi_j ^{-1} + \bm{H_jH_j}')^{-1}\bm H_j)\bm z_j'$

\textbf{Factor Loadings:}
We assume a lower triangular loadings matrix with a positive diagonal; the extension to other constraints is straightforward.
Let $k_j = \min(k, j)$ and $\bm H_j' $ be the $n\times k_j$ matrix with entries $\eta_{ik}$ for $1\leq k\leq k_j$ and $1\leq i \leq n$. Update nonzero elements in row $j$ of $\bm\Lambda$ as
$
\bm \lambda_j'\sim N({\bm{\hat \lambda_j}}'/v_j,\ (\bm\Psi_j^{-1} + \bm{H_jH_j}')^{-1})
$
where $\hat{\lambda_j}' = (\bm\Psi_j ^{-1} + \bm{H_jH_j}')^{-1}\bm{H_jz_j}'$ and $\lambda_{jj}$ is restricted to be positive if $j\leq k$.

\textbf{Hyperparameters:} Update $(1/\psi_{jh}|-)\sim InvGauss(|\xi_{jh}/\lambda_{jh}|,\ \xi_{jh}^2)$ and $(\xi_{jh}|-)\sim Ga(\alpha+1,\ \beta+|\lambda_{jh}|)$ where $InvGauss(a, b)$ is the inverse-Gaussian distribution with mean $a$ and scale $b$.

\textbf{Factor scores:} Draw $\bm\eta_i$ from 
$
(\bm\eta_i|-)\sim N\left(\inv{[\bm\Lambda'\bm\Lambda + I]}\bm\Lambda'z_i,\inv{ [\bm\Lambda'\bm\Lambda  + I]} \right)
$

\textbf{Augmented Data:} Update $\bm Z$ elementwise from
\beq
(z_{ij} | -) \sim  TN\left(\sumi{k}\lambda_{jh}\eta_{ki}, 1, z_{ij}^l , z_{ij}^u\right)\label{eq:zfc}
\eeq
where $TN(m, v, a, b)$ denotes the univariate normal distribution with mean $m$ and variance $v$ truncated to $(a, b)$, $z_{ij}^l = \max\{z_{i'j} : y_{i'j} < y_{ij}\}$ and $z_{ij}^u = \min\{z_{i'j} : y_{i'j} > y_{ij}\}$. If $y_{ij}$ is missing then $(z_{ij} | - )\sim  N(\sumi{k}\lambda_{jh}\eta_{ki}, 1)$. Note that \eqref{eq:zfc} doesn't require a matrix inversion since $(z_{ij}\id z_{ij'}\ |\ \bm\Lambda, \bm\eta_i, \bm Y)$ for $j\neq j'$, a unique property of our factor analytic representation and a significant computational benefit as $p$ grows. 

The PX-Gibbs sampler has mixing behavior at least as good as Gibbs sampling under the original model (which fixes $\bm V = \bm I$) \citep{Liu1999, Meng1999}, and the additional computation is negligible. The PX-Gibbs sampler often increases the smallest effective sample size (associated with the largest loadings) by an order of magnitude or more in both real and synthetic data. The improved mixing is also vital for the multimodal posteriors sometimes induced by shrinkage priors. To our knowledge this is the first application of PX to factor analysis of mixed data, but PX has previously been applied to Gibbs sampling in Gaussian factor models by \cite{dunson-default-fa} who introduce scale parameters for $\bm\eta_i$ to reduce dependence between $\bm H$ and $\bm\Lambda$. Since MCMC in our model suffers primarily from dependence between $\bm Z$ and $\bm \Lambda$ our approach is more appropriate. \cite{hoff-erl} and \cite{Dobra2011} also use  priors on unidentified covariance matrices to induce a prior on correlation matrices in Gaussian copula models. But the motivation there is simply to derive tractable MCMC updates and dependence between the priors on $\bm C$ and $\bm V$ precludes our strategy of choosing an optimal PX prior, limiting the benefits of PX. 

\subsection{Posterior Inference}

Given MCMC samples we can address a number of inferential problems. The posterior distribution of the factor scores $\bm\eta_{i}$ provide a measure of the latent variables for each data point, describing a projection of the observed data into the latent factor space, and the factors themselves are characterized by the variables which load highly on them. Even if the factors are not directly interpretable this is a very useful exploratory technique for mixed data which is robust to outliers and handles missing data automatically (unlike common alternatives such as principal component analysis). 

We can also do inference on conditional or marginal dependence relationships in $\bm y_i$. Here there is no need for identifying constraints in $\bm\Lambda$ which simplifies model specification. Tests of independence like $H_0: c_{jj'}\leq\epsilon$ versus $H_1: c_{jj'}>\epsilon$ are simple to construct from MCMC output. When the variables are continuous the conditional dependence relationships are encoded in $\bm R = \bm C^{-1}$ which we can compute as
\beq
\bm R = (\bm{\tilde\Lambda}\bm{\tilde\Lambda}'+\bm U)^{-1} = 
\bm U^{-1}-\bm U^{-1}\bm{\tilde\Lambda}
[\bm I + \bm{\tilde\Lambda}'\bm U^{-1}\bm{\tilde\Lambda}]^{-1}
\bm{\tilde\Lambda}'\bm U^{-1}\label{eq:factorcoef}
\eeq
Eq. \eqref{eq:factorcoef} requires calculating only $k$-dimensional inverses, rather than $p$-dimensional inverses, a significant benefit of our factor-analytic representation. 

As discussed in Section \ref{section:gausscop} the presence of discrete variables complicates inference on conditional dependence. Additionally, two  discrete variables may be effectively marginally independent even if $|c_{jj'}|>0$ simply by virtue of their levels of discretization. For these reasons, and for more readily interpretable results, it can be valuable to consider aspects of the posterior predictive distribution $\pi(\bm y^*|\bm Y)$. Under our semiparametric model this distribution is somewhat ill-defined, but we can sample from an approximation to $\pi(\bm y^*|\bm Y)$ by drawing $\bm{\tilde\Lambda}$ via the PX-Gibbs sampler, drawing $\bm z^*\sim N(0, \bm{\tilde\Lambda}\bm{\tilde\Lambda}'+ \bm U)$ and setting $y^*_j = \hat F^{-1}_j(\Phi(z^*_j))$ where $\hat F_j$ are estimators of each marginal distribution. This disregards some uncertainty when making predictions; \cite{hoff-erl} provides an alternative based on the values of $z_{ij}^l,\ z_{ij}^u$ from \eqref{eq:zfc} but (in keeping with his observations) we find both approaches to perform similarly.

Posterior predictive sampling of conditional distributions is detailed in Section 2 of the supplement. Importantly, the factor-analytic representation of $\bm C$ allows us to directly sample any conditional distributions of interest (rather than using rejection sampling from the joint posterior predictive) by reducing the problem of sampling a truncated multivariate normal distribution to that of sampling independent truncated univariate normals. 
 
\section{Simulation Study}\label{section:simstudy}

When fitting models in the following simulations we used the $GDP(3,1)$ prior for $\lambda_{jk}$ and take $1/\sigma^2\sim Gamma(2,2)$ for the Gaussian factor model and uniform priors on the cutpoints in the probit model. The cutpoints in the probit model were updated using independence Metropolis-Hastings steps with a proposal derived from the empirical cdfs. All models were fit using our R package.

\subsection{Relative efficiency}\label{section:sim-eff}
First we examine finite-sample relative efficiency of the extended rank likelihood in the ``worst-casse scenario" (for our method). We compare the posterior mean correlation matrix under the Gaussian copula factor model with the extended rank likelihood to that under 1) a Gaussian factor model, when the factor model is true and 2) a probit factor model, when the probit model is true. Both are special cases of the Gaussian copula factor model so we can directly compare the parameters.

\begin{figure}
\begin{center}
\includegraphics[width=0.95\textwidth]{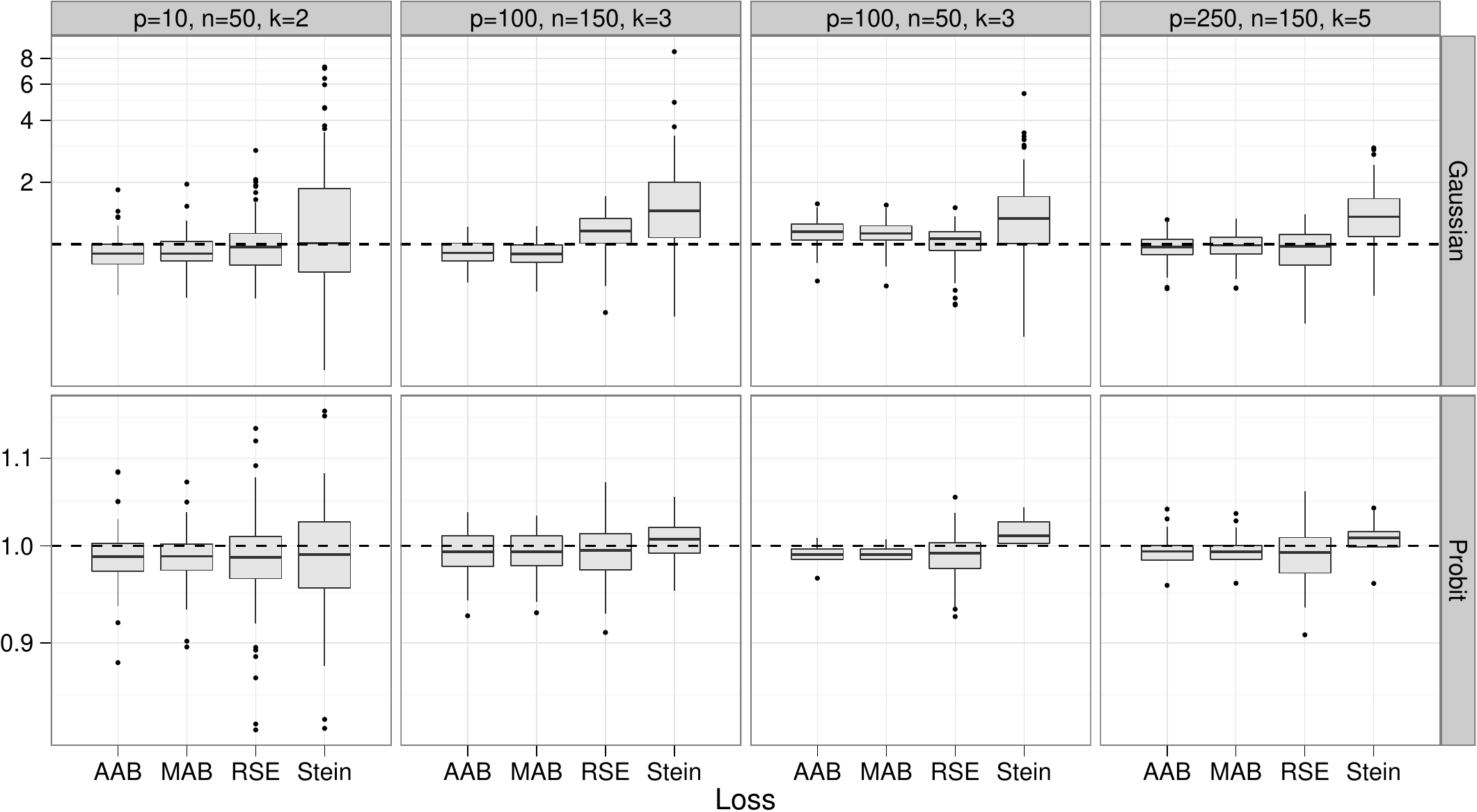}
\end{center}
\caption{Efficiency (ratio of the loss under our model to that under the Gaussian/probit model) of the posterior mean under a range of loss functions}
\label{fig:simresult}
\end{figure}

The true (unscaled) factor loadings were sampled iid $GDP(3,1)$. For the probit case each margin had five levels with probabilities sampled $Dirichlet(1/2, \dots, 1/2)$. We fix $k$ at the truth; additional simulations suggest that the relative performance is similar under misspecified $k$.  
We performed 100 replicates for various $p/k/n$ combinations in Fig. \ref{fig:simresult}. Each model was fit using 100,000 MCMC iterations after a 10,000 burn-in, keeping every 20th sample. MCMC diagnostics for a sample of the fitted models indicated no convergence issues. We assess the performance of each method by computing a range of loss functions: Average and maximum absolute bias ( $\frac{2}{p(p-1)}\sum_{i<j<p} |\hat c_{ij} -  c_{ij}|$ and $\max_{i<j<p}|\hat c_{ij} -  c_{ij}|$ respectively), root squared error: $\left[2\sum_{i<j<p} (\hat c_{ij} -  c_{ij})^2\right]^{1/2}$ and Stein's loss: $\tr(\hat{\bm C}\bm C^{-1})+\log\det(\hat{\bm C}\bm C^{-1})-p$. Stein's loss is (up to a constant) the KL divergence from the Gaussian copula density with correlation matrix $\bm C$ to the Gaussian copula density with correlation matrix $\hat{\bm{C}}$ and is therefore natural to consider here.

Figure \ref{fig:simresult} shows that the two methods are more or less indistinguishable in the probit case. Our method slightly outperforms the probit model in many cases because we do not have to specify a prior for the cutpoints. There are also some computational benefits here since in the copula model we avoid Metropolis-Hastings steps for the marginal distributions. In the continuous case our model also does well, although the Gaussian model is somtimes substantially more efficient under Stein's loss. But as $p$ grows our model is increasingly competitive. 

\subsection{Misspecification Bias and Consistency}\label{section:sim-bias}

The previous simulations suggest that the loss in efficiency in worst-case scenarios is quite often minimal. To illustrate the practical benefit of our model (and the impact of our consistency result) in a realistic scenario we simulated data from a one-factor Gaussian copula factor model using the marginal distributions from Section \ref{section:perisk} (Fig. \ref{periskhist}). For simplicity we take $\bm{\tilde\Lambda} = \tilde\lambda \bm{1}$ and consider $\tilde\lambda=0.7$ and $0.8$ (although we did not constrain the loadings to be equal when fitting models to the simulated data). Results in Section \ref{section:perisk} suggest that these are plausible values.

\begin{figure}
\begin{center}
\includegraphics[width=0.95\textwidth]{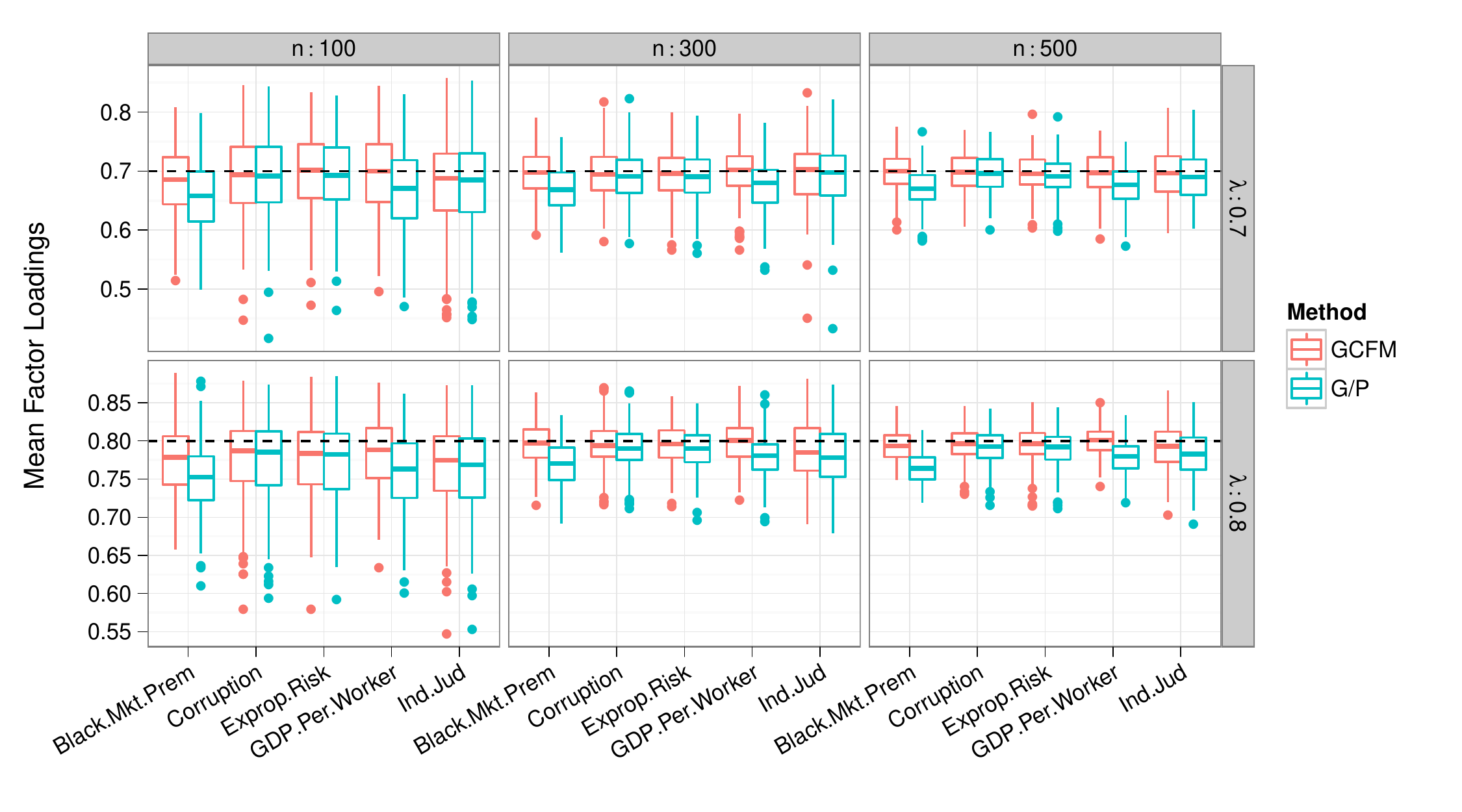}
\end{center}
\caption{Posterior mean factor loadings using 100 simulated datasets generated with the margins in Section 5 using our model (GCFM) versus a mixed Gaussian/probit model (G/P).}
\label{fig:perisk-simresult}
\end{figure}

Figure \ref{fig:perisk-simresult} shows that factor loadings for the two continuous variables (Black.Mkt.Premium and GDP.Per.Worker) are underestimated by the Gaussian/probit model. When all the variables are dependent there is a ``ripple" effect so that even factor loadings for discrete variables are subject to some bias. We should expect this behavior in general -- the copula correlations bound the observed Pearson correlations from above (in absolute value), with the bounds obtained only under Gaussian margins. The difference between the Pearson and copula correlation parameters, and hence the asymptotic bias, depends entirely on the form of the marginal distributions. This makes proper choices of transformations critical in the parametric model.  Our model relieves this concern entirely.  Although the magnitude of these effects is relatively mild here there is little reason to suspect this is true in general, especially in more complex models with multiple factors and a larger number of observed variables. 

\section{Application: Political-Economic Risk}\label{section:perisk}

\begin{figure}
\begin{center}
\includegraphics{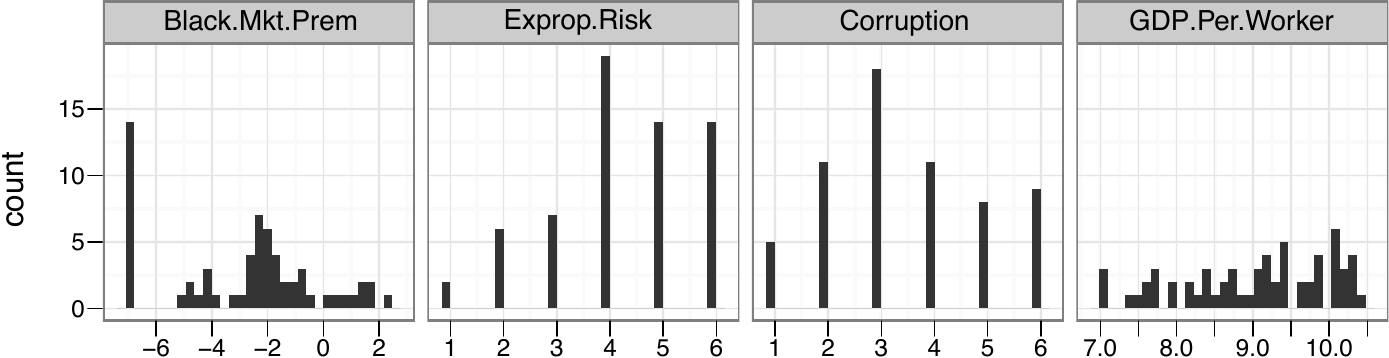}
\end{center}
\caption{Distributions of 4 variables from the political risk dataset in \cite{quinn}. The fifth, Ind.Jud, is binary with 34/62 ones.}
\label{periskhist}
\end{figure}

\cite{quinn} considers measuring political-economic risk, a latent quantity, using five proxy variables and a Gaussian/probit factor model. The author defines political-economic risk as the risk of a state ``manipulat[ing] economic rules to the advantage of itself and its constituents" following \citep[pp. 808]{nw1989} . The dataset includes five indicators recorded for 62 countries: independent judiciary,  black market premium, lack of appropriation risk, corruption, and gross domestic product per worker (GDPW) (Fig. \ref{periskhist}). Additional background on political-economic risk and on the variables in this dataset is provided by \cite{quinn}, and the data are available in the R package MCMCpack. \cite{quinn} transforms the positive continuous variables GDPW and black market premium by  $\log(x)$ and $\log(x+0.001)$ (resp.). The disproportionate number of zeros in black market premium (14/62 observations) leaves a large spike in the left tail and the normality assumption is obviously invalid. Since \cite{quinn} has already implicitly assumed a Gaussian copula, our model is a natural alternative to the misspecified Gaussian/probit model used there.

\begin{figure}
\begin{center}
\includegraphics[height=3in]{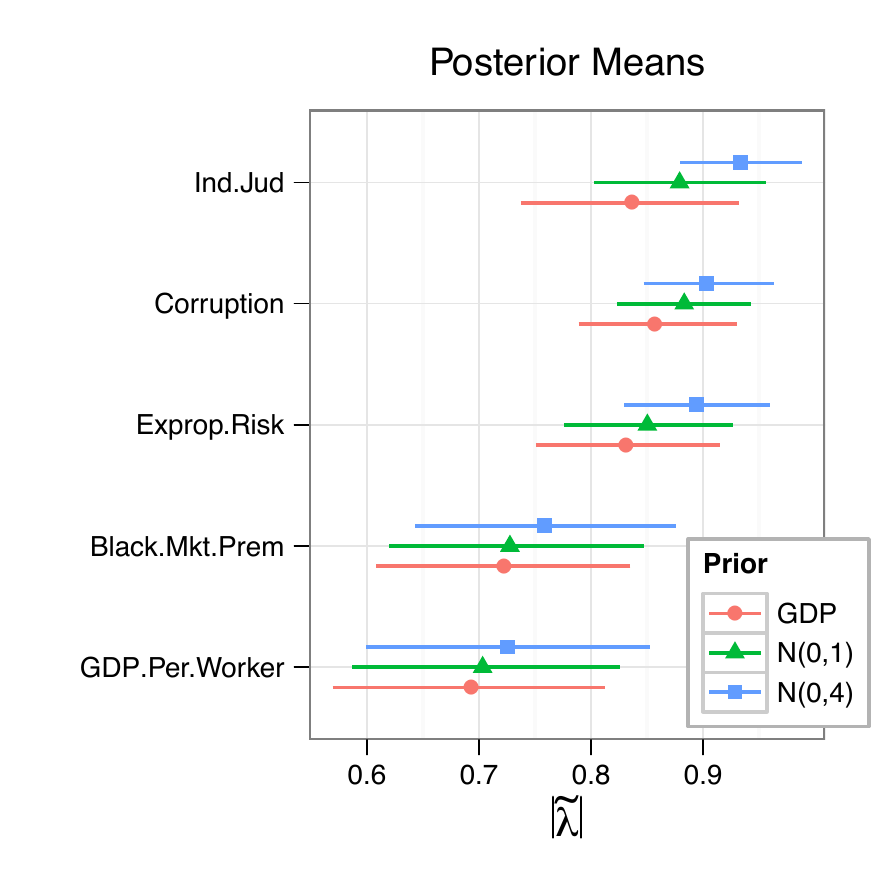}
\end{center}
\caption{Posterior means/90\% HPD intervals for scaled factor loadings under the different priors. Differences due to priors are larger for discrete variables, and largest for Ind.Jud (which is binary)}
\label{fig:perisk-priors}
\end{figure}

To explore sensitivity to prior distributions we fit the copula model under several priors: $GDP(3,1)$, $N(0,1)$ and the $N(0,4)$ priors used by \cite{quinn}. We use 100,000 MCMC iterations and save every $10^{th}$ sample after a burn-in of 10,000 iterations. Standard MCMC diagnostics gave no indication of lack of convergence. Figure \ref{fig:perisk-priors} shows posterior means and credible intervals for the scaled loadings under each prior. Note that the $N(0,4)$ prior, intended to be noninformative, is actually very informative on the scaled loadings (Fig \ref{fig:normuniq}). It pulls the scaled loadings toward $\pm 1$, with most pronounced influence on the binary variable Ind.Jud and the other categorical variables. The GDP prior instead shrinks toward zero as we would expect.

\begin{figure}
\begin{center}
\includegraphics[width=0.9\textwidth]{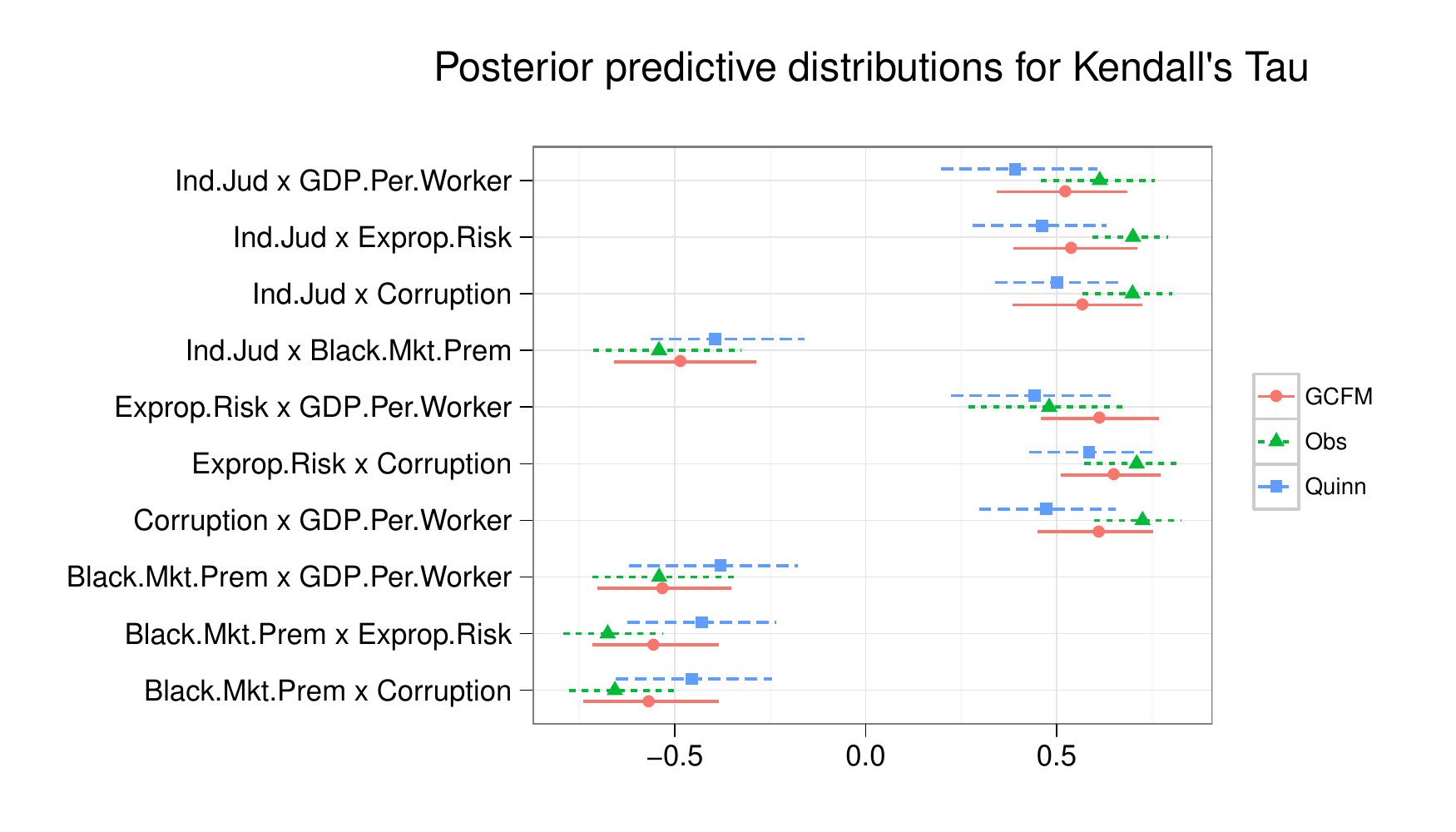}
\end{center}
\caption{Posterior predictive mean and 95\% HPD intervals of Kendall's $\tau$ under our model (Cop) and the Gaussian-probit model (Mix) as well as the observed values and bootstrapped 95\% confidence intervals.}
\label{fig:perisk-postpred-tau}
\end{figure}

We also compare our model to the Gaussian/probit model in \cite{quinn}, but using the $GDP(3,1)$ prior in both cases. Figure \ref{fig:perisk-postpred-tau} shows posterior predictive means and credible intervals for Kendall's tau, as well as the observed values and bootstrapped confidence intervals. Our model fits well, considering the limited sample size, and fits almost uniformly better than the Gaussian-probit model. Other posterior predictive checks on rank correlation measures and in subsets of the data show similar results. 

\begin{figure}
\begin{center}
\includegraphics{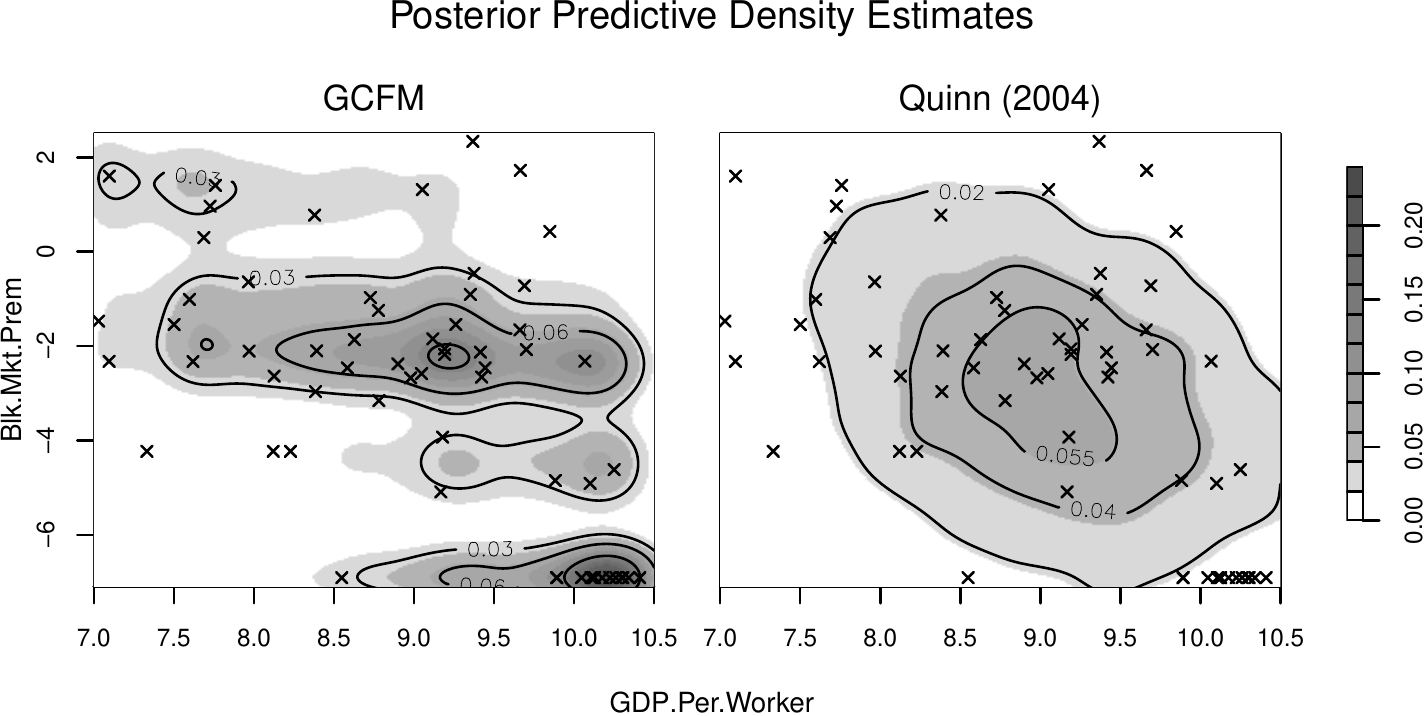}
\end{center}
\caption{Posterior predictive distributions of log GDP and log black market premium, with observed data scatterplots. Note the cluster of points in the bottom-right corner; even though they represent over 20\% of the sample the predictive density from the model in \cite{quinn} assigns very little mass to this area.}
\label{fig:perisk-postpred}
\end{figure}

\begin{figure}
\begin{center}
\includegraphics[scale=0.95]{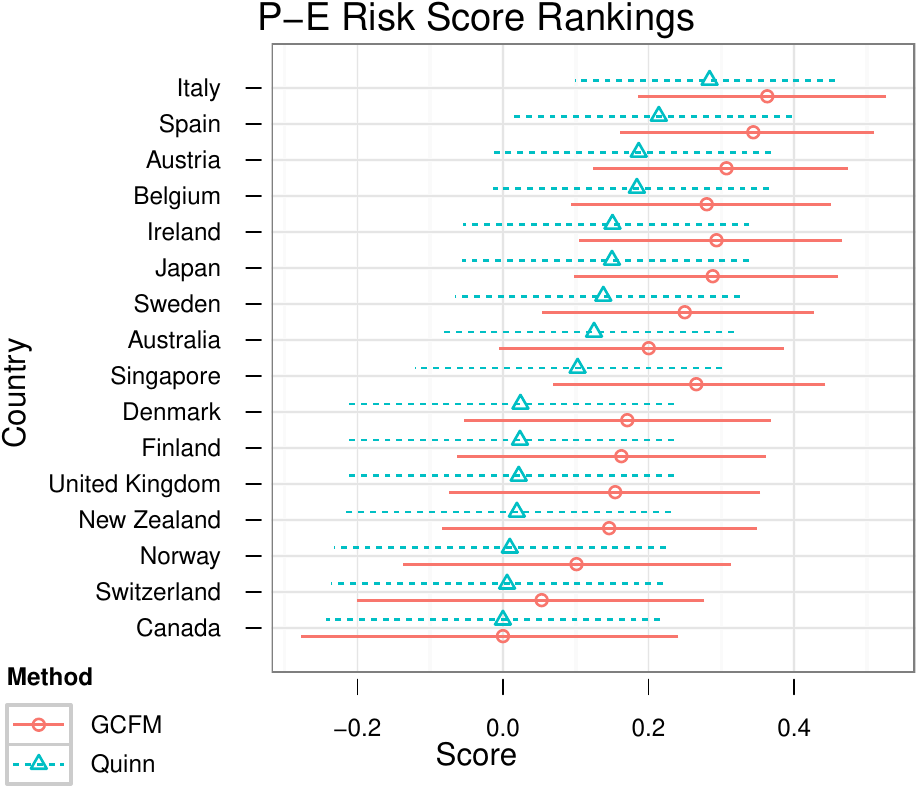}
\end{center}
\caption{Comparison of the political-economic risk ranking obtained via our model and the mixed-data factor analysis of \cite{quinn}. Points are posterior means and lines represent marginal 90\% credible intervals.}
\label{fig:perisk1}
\end{figure}

Incorrectly assuming a normal distribution for log black market premium is especially damaging. The copula correlation between GDPW and black market premium (on which the data are most informative) is underestimated in the Gaussian/probit model: mean $-0.33$ and 95\% HPD interval $(-0.46, -0.22)$ as opposed to $-0.56$ and $(-0.73, -0.40)$  under our model. This is also evident in the posterior predictive samples of Kendall's $\tau$ in Fig. \ref{fig:perisk-postpred-tau}. Figure \ref{fig:perisk-postpred} shows density estimates of draws from the bivariate posterior predictive of black market premium and GDPW. The Gaussian/probit model is clearly not a good fit, assigning very little mass to the bottom-right corner (which contains almost 25\% of the data). The Gaussian copula factor model assigns appropriately high density to this region. 
Estimates of the latent variables are impacted as well: Figure \ref{fig:perisk1} plots the mean factor scores from each model (after shifting and scaling to a common range) for low-risk countries. The seven countries with the lowest risk have identical covariate values except on GDPW. Our model infers mean scores that are sorted by GDPW (higher GDPW yielding a lower score). The Gaussian/probit model instead assigns these countries almost identical scores. 

\section{Discussion}\label{section:conclusion}

In this paper we have developed a new semiparametric approach to the factor analysis of mixed data which is both robust and efficient. We propose new default prior distributions for factor loadings which are more suited to routine use of this model (and similar models, such as probit factor models). As a byproduct we also induce attractive new priors on correlation matrices in Gaussian copula models; these are both more flexible and parsimonious than the inverse Wishart prior used by \cite{hoff-erl}, and much more efficient computationally than the graphical model priors of \cite{Dobra2011}.  They admit optimal parameter expansion schemes which are easy to implement, and are readily extended to informative specifications, to include covariates or to more complex latent variable models. 

We have not considered the issue of uncertainty in the number of factors, but it is straightforward to do so by adapting existing methods for Gaussian factor models.  In addition to posterior predictive checks, these include stochastic search \citep{sparse-fm}, reversible jump MCMC \citep{Lopes2004b}, Bayes factors \citep{dunson-default-fa, Lopes2004b} and nonparametric priors \citep{Paisley2009, Bhattacharya2011}. The latter are especially promising when interest lies in $\bm C$ since they preserve the computational advantages of factor-analytic priors while providing full support on correlation matrices (which fails for fixed $k<p$). Particularly when the plausible range of $k$ is quite small, posterior predictive checking can be very effective.


\appendix

\section{Proof of Theorem 1}\label{appendix:proof}
\begin{proof}
We require a variant of Doob's theorem, presented in \cite{Gu2009}:

\begin{doob}
Let $X_i$ be observations whose distributions depend on a parameter $\theta$, both taking values in Polish spaces. Assume $\theta\sim \Pi$ and $X_i|\theta\sim P_\theta$. Let $\mathcal{X}_N$ be the $\sigma$-field generated by $X_1, \dots, X_N$ and $\mathcal{X}_\infty = \sigma\langle \bigcup_{i=1}^\infty \mathcal{X}_i\rangle$. If there exists a $\mathcal{X}_\infty$ measurable function $f$ such that for $(\omega, \theta)\in \Omega^\infty\times\Theta$,  $\theta = f(\omega)\ \text{a.e.}\ [P_\theta^\infty\times\Pi]$ 
then the posterior is strongly consistent at $\theta$ for almost every $\theta$ $[\Pi]$.
\end{doob}

Therefore we must establish the existence of a consistent estimator of $\bm{C}$ which is measurable with respect to the $\sigma$-field generated by the sequence $\{D(\bm Y^{(m)})\}_{m=1}^\infty$ (a coarsening of the $\sigma$-field generated by $\{Y^{(m)}\}_{m=1}^\infty$).
Let $R_{nij} = \sum_{h=1}^n \ind{[y_{hj}\leq y_{ij}]} = n\hat F_{j}(y_{ij})$. Let $\mathbf{R}_{ni}(\bm Y^{(n)})$ be the $p$-vector with entry $j$ given by $R_{nij}$  and let $\mathbf{R}_n(\bm Y^{(n)}) = \{\mathbf{R}_{ni}\}_{i=1}^n$. Observe that the information contained in the extended rank likelihood (namely the boundary conditions in the definition of the set $D(\bm Y^{(n)})$) is equivalent to the information contained in $\bm R_n(\bm Y^{(n)})$. Hence a function that is measurable with respect to $\mathcal{R}_n$, the $\sigma$-field generated by $\{\bm R_m(\bm Y^{(m)})\}_{m=1}^n$, is also measurable with respect to the $\sigma$-field generated by $\{D(\bm Y^{(m)}\}_{m=1}^N$ and we may work exclusively with the former. 

Let $\hat U_{nij} = \frac{R_{nij}}{n+1}$ and $\bm{\hat{U}}_{ni} = (\hat U_{ni1}, \dots \hat U_{nip})'$. Then $\hat U_{nij} \wpone U_{ij}$ where $U_{ij} = F_j(y_{ij})$ by the SLLN, so $\bm{\hat{U}}_{ni} \wpone \bm U_i$ and therefore $\bm U_i$ is $\mathcal{R}_\infty = \sigma\langle \bigcup_{i=1}^\infty \mathcal{R}_i \rangle$ measurable. Note that if $F_j$ is discrete $U_{ij}$ is merely a relabeling of $y_{ij}$ (each category/integer is ``labeled" with its marginal cumulative probability). So $\bm U_i$ is a sample from a Gaussian copula model with correlation matrix $\bm C_0$ where the continuous margins are all $U[0,1]$ and the discrete marginal distributions are completely specified. The problem of estimating $\bm C$ from $\bm U_i$ reduces to estimating ordinary and polychoric/polyserial correlations with fixed marginals and it is straightforward to verify that the distribution of $\bm U_i$ is a regular parametric family admitting a consistent estimator of $\bm C$, say $h_N(\bm U_1, \dots, \bm U_N)$. 
Therefore there exists a sequence of $\mathcal{R}_\infty$ measurable functions $h_N(\bm U_1, \dots, \bm U_N)\rightarrow h(\bm U_1, \bm U_2, ...)=\bm C_0$ almost surely and
\begin{align}
\bm C_0 &= h(\bm U_1, \bm U_2, ...)= h^*(\{ \mathbf{R}_{Ni} : N\geq 1, 1\leq i\leq N \})\ \text{a.s.}\ [G^\infty_{\bm C_0, F_1, \dots, F_p}]\label{eq:estimator}
\end{align}
where \eqref{eq:estimator} holds because a null set under the measure induced by $\bm R_n(\bm Y^{(n)})$ is also null under $G^\infty_{\bm C_0, F_1, \dots, F_p}$.
\end{proof}

\section{Validity of the PX Sampler}\label{appendix:px}

Let $\bm\Theta$ be the inferential parameters and let $s_j=\bm{z_j}(\bm I - \bm H_j'(\bm\Psi_j ^{-1} + \bm{H_jH_j}')^{-1}\bm H_j)\bm z_j'$
Our working prior for $(v_{1},\dots v_{p})$ is $\prod_{j=1}^P IG(v^2_j; n_0/2, n_0/2)$. To verify that samples of $\bm\Theta$ from the PX-Gibbs sampler have stationary distribution $\pi(\bm\Theta|\bm Y)$ we need to show that as $n_0\rightarrow 0$ the transition kernels under the marginal sampling scheme (alternately drawing from $\pi(\bm{W|\Theta, Y})$ and $\pi(\bm{\Theta|W})$) and the blocked  sampling scheme (alternately drawing from $\pi(\bm{W|\Theta, V, Y})$ and $\pi(\bm{\Theta, V|W})$) converge \citep{Meng1999}.
The $t^{th}$ updates under the two schemes are as follows:

\noindent\textbf{Scheme 1:} Draw $1/v_{0j}^2 \sim Ga(n_0/2, n_0/2)$ and $1/v_{1j}^{2} \sim Ga\left(\frac{n_0+n}{2}, 
\frac{n_0 + v_{0j}^2s_j}{2}\right)$. Set $r = v_{j0} / v_{j1}$ and draw ${\bm \lambda_j} \sim N(r \bm{\hat{\lambda_j}}', (\bm \Psi_j^{-1} + \bm{HH}')^{-1})$ 

\noindent\textbf{Scheme 2:} Draw $1/v_{tj}^2 \sim Ga\left(\frac{n_0+n}{2}, 
\frac{n_0 + v_{(t-1)j}s_j}{2}\right)$. Set $r = v_{(t-1)j} / v_{tj} $ and draw ${\bm \lambda_j} \sim N(r \bm{\hat{\lambda_j}}', (\bm \Psi_j^{-1} + \bm{HH}')^{-1})$ 

Updates for the rest of $\bm\Theta$ under both schemes are the same as in Section \ref{section:gibbs}. As $n_0\rightarrow 0$ under Scheme 1 the distribution of $1/v_{0j}^{2}$ approaches a point mass at 1 and Scheme 1 converges to Scheme 2 with $n_0=0$.

\section{Supplementary Material}

\subsection{Conditional independence}\label{appendix:ci}
Assume $F(Y_1, Y_2, Y_3)$ has a Gaussian copula with correlation matrix $\bm C$, that $Y_3$ is discrete, and that $r_{12}=0$. Let $(Z_1, Z_2, Z_3)\sim N(\bm 0, \bm C)$ and $\mathcal{B}_c = (F_3(c-1),\ F_3(c)]$ for $c$ in the domain of $Y_3$ and define $g_j(z_3)=\Phi\left(F_j(y_j) - c_{j3}z_3)/(1-c_{j3}^2)^{1/2}\right)$. It is straightforward to show that
\begin{align}
Pr(Y_1\leq y_1|Y_3=c)Pr(Y_2\leq y_2|Y_3=c)
&=E(g_1(z_{3}))E(g_2(z_{3}))\label{eq:condobs}\\
Pr(Y_1\leq y_1,\ Y_2\leq y_2\ |\ Y_3=c)
&=E(g_1(z_{3})g_2(z_{3}))\label{eq:condlatent}
\end{align}
where the expectations are with respect to $\pi(z_3|y_3=c) = TN(0, 1, F_3(c-1), F_3(c))$ and \eqref{eq:condlatent} holds because $\pi(z_1, z_2|z_3)=\pi(z_1|z_3)\pi(z_2|z_3)$ when $r_{12}=0$. Since $g_1, g_2$ are monotone it is well known that $E(g_1(z_{3})g_2(z_{3}))\neq E(g_1(z_{3}))E(g_2(z_{3}))$ (and $Y_1, Y_2$ are dependent given $Y_3$) unless one or both functions are a.s. constant, which occurs only if one or both of $Y_1, Y_2$ are marginally independent of $Y_3$ ($c_{13}c_{23}=0$). This result extends to conditioning on one discrete variable and any number of continuous variables since conditioning on a continuous variable $Y_4=y_4$ implies that $Pr(z_4=\Phi^{-1}(F(y_4))=1$, and $\pi(z_3|y_3, z_4)$ is again univariate truncated normal (with a different mean and variance).

\subsection{Posterior Predictive Conditional Distributions}\label{appendix:conditionals}

To sample from conditional posterior predictive distributions such as $\pi(y_1^*\ |\ \bm y_{(-1)}^*=\bm x, \bm Y)$ we could sample from $\pi(\bm y^*|\bm Y)$ and discard draws where $y^*_j\neq x_j$ for any $2\leq j\leq p$. This approach can be wasteful computationally since even in moderate dimensions most samples will be discarded. Instead we might prefer to estimate this distribution directly.
We can write $Pr(y^*_{1}\leq y\ |\ \bm y^*_{(-1)}=\bm x,\ \bm Y)$ as
\beq
\int\limits_{\mathcal{C}}\int\limits_{{\mathbb{R}^{p-1}}}
\left(\int\limits_{-\infty}^{{\hat{F}_1(y)}}
\pi(z_1^*|\ \bm z^*_{(-1)}, \bm C)\,dz_1^*\right)
\pi(\bm z^*_{(-1)}|\ \bm y^*_{(-1)}=\bm x, \bm C)
\pi(\bm C\ |\ \bm Y)\,d\bm z^*_{(-1)}\,d\bm C
\label{eq:zpp}
\eeq
Assume that $y_2,\dots y_p$ are discrete, or that the empirical cdfs are used for $\hat F_j$ (if $y_{j}$ is continuous and $\hat F_j$ is a smooth estimator then $z^*_{j}=\Phi^{-1}(\hat F(x_{j}))$ is fixed in the following). Then $\pi(\bm z^*_{(-1)}\ |\ \bm y^*_{(-1)}=\bm x, \bm C)$ is the $(p-1)$-dimensional truncated normal distribution $N(\bm 0, \bm C_{(-1)})$ where $\bm C_{(-1)}$ is obtained by dropping the first row and column of $\bm C$,  restricted to the set 
$\mathcal{B}_x =\{ \bm z^*_{(-1)};\ 
\Phi^{-1}(\hat F_j(x_j^-))<z^*_j\leq \Phi^{-1}(\hat F_j(x_j))\ 
\forall\ 2\leq j\leq p\}$ (where $F(x^-)$ is the lower limit of $F$ at $x$). To estimate \eqref{eq:zpp} from MCMC output we need to draw from this distribution (at least) once for every sample of $\bm{C}$.
For a general $\bm C$ this is prohibitive unless $p$ is very small, but our factor-analytic representation allows us to efficiently draw from $\pi(\bm z^*_{(-1)}\ |\ \bm y_{(-1)}=\bm x, \bm C)$ by sampling $(p-1)$ \emph{univariate} truncated normals: Let $\bm{\tilde\Lambda}_{(-1)}$ be $\bm{\tilde\Lambda}$ with the first row removed and $\bm U_{(-1)}$ be $\bm U$ with the first row and column removed. Since $\bm C_{(-1)} = \bm{\tilde\Lambda}_{(-1)}\bm{\tilde\Lambda}_{(-1)}'+\bm U_{(-1)}$ we have
\begin{align*}
\pi(\bm z^*_{(-1)}\ |\ \bm y^*_{(-1)}=\bm x, \bm{\tilde\Lambda}_{(-1)})
&\propto N(\bm z^*_{(-1)};\ \bm 0, \bm{\tilde\Lambda}_{(-1)}\bm{\tilde\Lambda}_{(-1)}'+\bm U_{(-1)})\ind{(z^*_{(-1)}\in \mathcal{B}_x)}\\
&\propto \int_{\mathbb{R}^k} \prod_{j=2}^p 
\left(TN(\bm{\tilde\lambda}_j\bm\eta, u_j, a_j, b_j)\right)
N(\bm\eta; \bm 0, \bm I)\,d\bm\eta
\end{align*}
where $a_j= \Phi^{-1}(\hat F_j(x_j^-))$, $b_j=\Phi^{-1}(\hat F_j(x_j))$ and $\bm\eta$ is an auxiliary variable. Therefore we can approximate \eqref{eq:zpp} as follows:

\ben
\item Draw $\bm{\tilde\Lambda}$ via the PX-Gibbs sampler, and draw $\bm\eta\sim N(\bm 0, \bm I)$
\item Draw $z^*_j\sim TN(\bm{\tilde\lambda}_j\bm\eta, u_j, a_j, b_j)$ for $2\leq j\leq p$
\item For each distinct value of $\bm y_1$ set $\tilde F^{(t)}(y_i) = \int_{-\infty}^{{\hat{F}_1(y_i)}}N(z_1^*;m,v)\,dz_1^*$
where
\begin{align}
m &= \bm{\tilde\lambda}_{1}\bm{\tilde\Lambda}'_{(-1)}[\bm{\tilde\Lambda}_{(-1)}\bm{\tilde\Lambda}_{(-1)}'+\bm U_{(-1)}]^{-1}\bm z^*_{(-1)}\nonumber\\
v &= 1-\bm{\tilde\lambda}_1\bm{\tilde\Lambda}_{(-1)}'[\bm{\tilde\Lambda}_{(-1)}\bm{\tilde\Lambda}_{(-1)}'+\bm U_{(-1)}]^{-1}\bm{\tilde\Lambda}_{(-1)}\bm{\tilde\lambda}_1'\label{eq:condstat}
\end{align}
\een
where again the matrix inverses in \eqref{eq:condstat} can be computed efficiently as in \eqref{eq:factorcoef}. This procedure provides estimates of the conditional cdf at the observed data points. For a discrete response we can then directly compute conditional probabilities, odds ratios, and so on. When $y_1$ is continuous these can be interpolated to give a histogram estimate of $\pi(y_1|\bm y_{(-1)}=\bm x)$ with support on the range of the observed data.  A number of modifications to this approach are possible; for example, to condition on a subset of $\bm y_{(-1)}$ we simply drop the irrelevant rows of $\bm\Lambda_{(-1)}$ and only perform step 3 for the $j^{th}$ variable if we are conditioning on $y_j$. 

This is a natural extension of factor regression models which posit a Gaussian factor model for $(y_i, \bm x_i')'$, implying a linear regression model for $\pi(y_i\ |\ \bm x_i)$ \citep{West2003, sparse-fm}. These are especially useful when $p>n$ as a model-based form of reduced rank regression (automatically selecting batches of correlated predictors by loading them highly on the same factor), or when there is missing data in $\bm X$. Here we have a flexible joint model which accommodates any ordered response or covariates while retaining the computational simplicity of factor regression models.

\bibliographystyle{apalike}
\bibliography{refs}

\end{document}